\documentclass[]{aastex631}
\usepackage[toc,page]{appendix}
\usepackage[caption = false, position=top]{subfig}
\usepackage{amsmath}
\usepackage{amssymb}
\usepackage[normalem]{ulem}
\usepackage{soul}
\usepackage{booktabs}
\usepackage{sidecap}
\usepackage{floatrow}
\usepackage{mathtools}
\usepackage{tgcursor}
\usepackage{url}
\def\UrlBreaks{\do\/\do-}

\shorttitle{Magnetic flux in the Sun emerges unaffected by supergranular-scale surface flows}
\shortauthors{Mani et al.}

\begin{document} 
\title{Magnetic flux in the Sun emerges unaffected by supergranular-scale surface flows}

\author[0000-0002-8707-201X]{Prasad Mani}
\affiliation{Department of Astronomy and Astrophysics, Tata Institute of Fundamental Research, Mumbai, India}
\email{prasad.subramanian@tifr.res.in}

\author[0000-0003-2536-9421]{Chris S. Hanson}
\affiliation{Center for Astrophysics and Space Science, NYUAD Institute, New York University Abu Dhabi, Abu Dhabi, UAE}

\author[0000-0001-8699-3952]{Siddharth Dhanpal}
\affiliation{Department of Astronomy and Astrophysics, Tata Institute of Fundamental Research, Mumbai, India}

\author[0000-0003-2896-1471]{Shravan Hanasoge}
\affiliation{Department of Astronomy and Astrophysics, Tata Institute of Fundamental Research, Mumbai, India}
\affiliation{Center for Astrophysics and Space Science, NYUAD Institute, New York University Abu Dhabi, Abu Dhabi, UAE}

\author[0000-0003-0896-7972]{Srijan Bharati Das}
\affiliation{Department of Geosciences, Princeton University, Princeton, NJ, USA}

\author[0000-0001-5850-3119]{Matthias Rempel}
\affiliation{High Altitude Observatory, National Center for Atmospheric Research, Boulder, CO 80307, USA}

\begin{abstract}
Magnetic flux emergence from the convection zone into the photosphere and beyond is a critical component of the behaviour of large-scale solar magnetism. Flux rarely emerges amid field-free areas at the surface, but when it does, the interaction between magnetism and plasma flows can be reliably explored. Prior ensemble studies identified weak flows forming near emergence locations, but the low signal-to-noise required averaging over the entire dataset, erasing information about variation across the sample. Here, we apply deep learning to achieve improved signal-to-noise, enabling a case-by-case study. We find that these associated flows are dissimilar across instances of emergence and also occur frequently in the quiet convective background. Our analysis suggests diminished influence of supergranular-scale convective flows and magnetic buoyancy on flux rise. Consistent with numerical evidence, we speculate that small-scale surface turbulence  and / or deep-convective processes play an outsize role in driving flux emergence.  
\end{abstract}

\section{Introduction}\label{section_introduction}
Active regions (ARs) are spatially and temporally extensive magnetic phenomena, extending from the solar interior to the corona, with a lifetime marked by formation, emergence, and eventual decay \citep{vanDriel2015} through dispersion of magnetic elements \citep{strous1996,schunker_2019, schunker2020} into the background turbulent convective field. Most ARs play host to sunspots \citep{rempel2011,rempel2012} - magnetic features characterized by evolving umbrae, penumbrae, and fine-structures, with high field strengths \citep[few $\sim$kG,][]{siu2019}. The dynamics of tangled magnetic field lines in morphologically complex active regions and sunspots underpin high-energy eruptive events such as flares \citep{toriumi2019} and coronal mass ejections \citep{webb2012}. A detailed inquiry into large-scale solar magnetism, vis-a-vis ARs, has indirect implications for understanding space-weather \citep{temmer2021}. 

ARs are hypothesized to be the surface manifestations of thin magnetic flux tubes \citep{cheung2014} generated in the interior \citep{charbonneau2020}, which then rise up \citep{birch2016} to the surface and above through a collection of processes termed {\it emergence}. The precise location of the dynamo is contested, with suggestions ranging from the base of the convection zone \citep{spiegel1980} to the near-surface layers \citep{brandenburg2005}. A comprehensive understanding of solar magnetism warrants that ARs be studied over the full domain - from birth to decay. Their associated flows have drawn attention \citep{gizon_2001,komm2008,hindman_2009} since magnetic fields and solar convection are thought to be intertwined \citep{stein2012lrsp}. A medley of findings, courtesy of individual emerging-active-region (EAR) studies \citep{kosovichev2009,zharkov2008,komm2008,hartlep2011}, fail to form a coherent picture of flux-emergence physics. This has motivated ensemble studies of isolated ARs, which report near-surface flows forming around an averaged EAR several hours prior to emergence. Chiefly, precursor-like horizontal convergent flows (inflows) in the vicinity of EARs \citep{martin-belda2017,loptein2017,birch_2019,braun2019,gottschling_2021} are commonly found to be correlated with emerging flux. Other studies show rotating magnetic features \citep{snodgrass1983,howard1992,kutsenko2021} and circulating flows near ARs \citep{hindman_2009,komm_2012}. 

Ensemble studies aim to mitigate the strong supergranular background flow \citep[$\sim300$ m/s,][]{rieutord_rincon_2010} by averaging over many EARs, in order to examine the weaker precursor flows \citep[$\sim40-50$ m/s,][]{birch_2019,gottschling_2021}. However, when undertaking ensemble studies, one must ensure minimal variance in the different properties of ARs such as its field morphology, surface area, net flux content, and the solar cycle to which it belongs. The latter requires accounting for Hale's law \citep{hale_1925} - magnetic polarities in the two hemispheres are statistically opposite in sign, and Joy's law \citep{hale_1919} - the leading polarity is tilted towards the equator in both the hemispheres. Target selection thus plays a vital role in the proper interpretation of results. The LBB survey \citep{leka2013,birch2013,barnes2014} is a helioseismology program that documents the properties of over 100 EARs - emergence time, location, and area, and investigates their subsurface dynamics. We use the SDO/HEAR (Helioseismic Emerging Active Region) survey of \citet{schunker_2016} and \citet{schunker_2019} that improved upon LBB survey and obtained close to 180 ARs for the time period 2010-2014. The averaging process in ensemble studies sacrifices crucial information about individual EARs in favor of suppressing background noise. Traditional investigations may well benefit from novel deep-learning techniques that are able to analyze poor signal to noise (SNR) data with more finesse \citep{ml_snr1,ml_snr2}. We demonstrate using convolution neural networks that even in a dataset thought to be consistent and minimally variant, flows around different ARs behave substantially differently.  

\section{Data analysis}
Time-series of continuum intensity and magnetograms, which capture variations in surface brightness and line-of-sight (LOS) magnetic field, respectively, are obtained from the Helioseismic and Magnetic Imager on board the Solar Dynamics Observatory  \citep[SDO/HMI,][]{scherrer_etal_2012,schou2012}. The observed spatial map size is $32^\circ\times32^\circ$ ($\sim389\times389$~Mm$^2$), centered on the EAR location, with a duration of 54 hours at a cadence of 45s (4320 frames). The spatial resolution for intensity-cubes is $0.04^\circ=0.486$Mm, and for magnetogram-cubes is $0.08^\circ=0.972$Mm. Regions are tracked at Snodgrass rotation rate \citep{snodgrass1984} and Postel projected. 

From the list of ARs catalogued in the SDO/HEAR survey \citep{schunker_2016,schunker_2019}, we pick 115 that emerge amidst a relatively quiet-Sun region (i.e., $P\leq2$ in the survey - a number assigned by visual inspection of the active region and its surroundings; the lower the number, the lesser any pre-existing magnetic field). We use their definition of emergence time - first compute the maximum absolute flux (corrected for LOS projection) in a 36-hr window after NOAA records the first appearance of sunspot, and subsequently fix emergence time to correspond to when the flux reaches 10\% of the maximum in that 36-hr window. The emergence location is defined as the flux-weighted centre of the LOS magnetic field at the emergence time. With $x$ pointing prograde (in the direction of rotation), $y$ pointing solar north, the emergence location is $(x,y)=(0,0)$. The total observed time of 54 hours is split into 36 hours pre-emergence and 18 hours post-emergence. We select ARs that lie within $50^{\circ}$ of central meridian in order to avoid limb effects. Each AR has a unique flux history depending on its magnetic field properties and the amount of pre-existing field in the vicinity of emergence \citep{leka2013}. Thus, the proposed thresholds for emergence time and location are intended only to aid statistical studies as opposed to making strong claims about emergence physics.

\begin{figure}%[!htb]
    \includegraphics[scale=0.6]{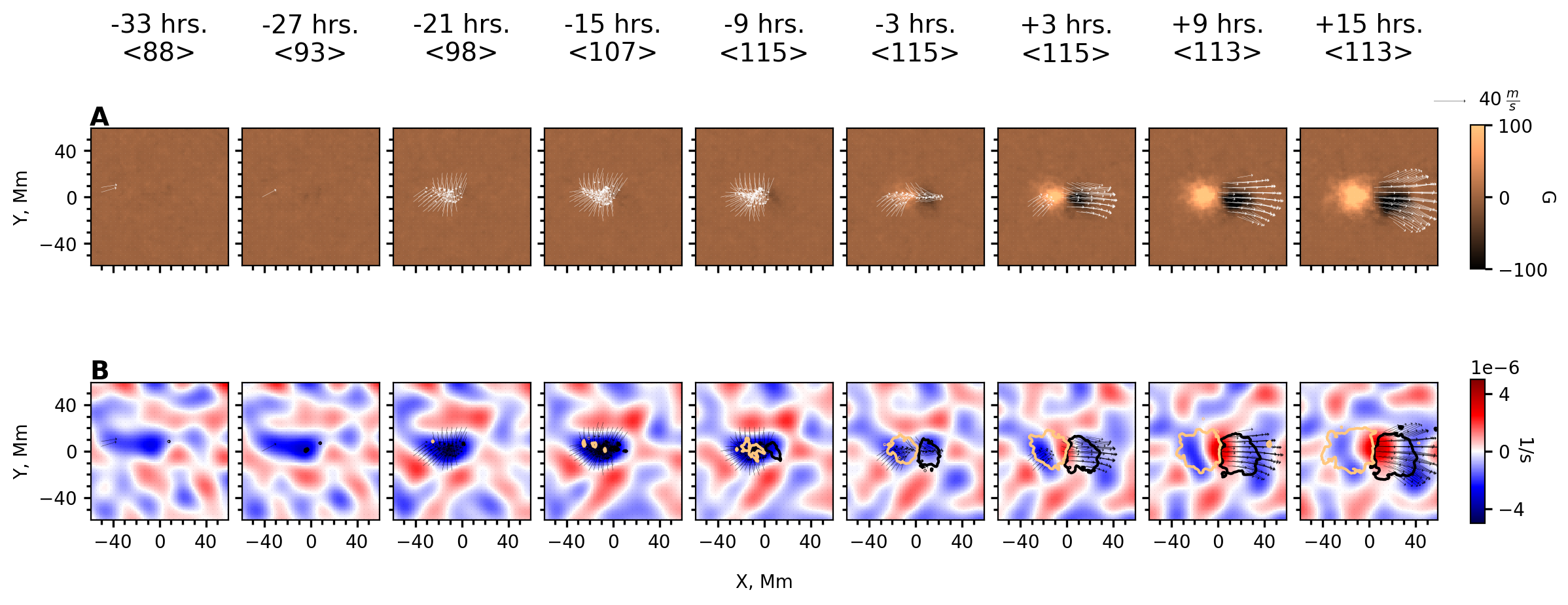}
    \caption{Ensemble average of all the ARs in the nine time intervals with \textbf{A}: LOS magnetograms, \textbf{B}: horizontal divergence ($\boldsymbol{\nabla}_h\cdot\bf{u}$). The mid-point of the time interval and the number of ARs averaged are stated in the title of each column. Magnetic-field contours of $\pm$10 G are overlaid on the flows to aid visualization. The first six columns belong to pre-emergence time ($t<0$), while the last three columns denote post-emergence time ($t>0$). {The white arrows in each panel denote velocity field of the plasma. Only speed$>$12 m/s are plotted. The top right of the figure denotes the arrow length corresponding to 40 m/s speed.}}
    \label{fig_main_fullset}
\end{figure}

The evolution timescales of ARs may vary from tens of hours to a few days, depending on their flux content \citep{vanDriel2015}. In order to better capture all the variations, the 54-hr observed duration is partitioned into nine contiguous 6-hr intervals and analyzed independently. Stipulating the central-meridian distance condition, we obtain a total of 88, 93, 98, 107, 115, 115, 115, 113, and 113 ARs for the nine intervals: $T_1=[-36,-30]$, $T_2=[-30,-24]$ $...$ $T_8=[6,12]$, $T_9=[12,18]$, with the numbers denoting hours from emergence time $t=0$. Ensemble-averaged LOS magnetic fields and horizontal-divergence of the flows ($\boldsymbol{\nabla_h}\cdot\bf{v}$, see equation ~\ref{lctdiv}) for the nine intervals are shown in Figure~\ref{fig_main_fullset}. Bipolar magnetic fields (A) steadily rise in strength as emergence time approaches (left to right in the figure). Our goal is to understand if the flows (B) drive / are driven by the emerging magnetic flux. To correctly interpret AR ensemble-study results, i.e., to identify whether a flow signal is correlated with flux emergence and not attributable to background noise, it must be compared against quiet-Sun flows. We build a deep-learning network that will predict the presence/absence of AR-like flow features in individual flow images with sufficient conviction.

\subsection{Data products}
Horizontal flows are obtained using Local Correlation Tracking \citep[LCT,][]{november1988} on the intensity-continuum data. LCT is an established method of inferring horizontal velocity field $[v_x, v_y]$ at the photosphere. The method examines the advection of convective granules \citep[$\sim$1~Mm, see][]{hathaway_2015} by underlying larger-scale flow systems (e.g., supergranules or EAR flows, $30-40$~Mm). Since granules are used as tracers, which are much smaller in size than supergranules, LCT is an effective method \citep[see][]{Rieutord01} to produce surface horizontal flows of supergranulation sizes, the length-scale in which we are interested. We use the \texttt{pyFLCT 0.2.2} (see Appendix~\ref{section_pyflct} and \citet{sunpy_community2020}) routine to obtain $[v_x, v_y]$ and get horizontal divergence $div$ and radial vorticity $curl$
% \begin{align}
    % div &= \partial_x v_x + \partial_y v_y,\label{lctdiv} \\
    % curl &= \partial_x v_y - \partial_y v_x.\label{lctcurl}
% \end{align}
\begin{equation}
    div = \partial_x v_x + \partial_y v_y,\label{lctdiv} 
\end{equation}
\begin{equation}
    curl = \partial_x v_y - \partial_y v_x.\label{lctcurl}
\end{equation}
Global-scale background flows on the Sun, such as differential rotation \citep{howe2009} and meridional circulation \citep{hanasoge2022}, can induce systematic variations in measured local flow velocities. This is undesirable as we are only interested in flow structures associated with AR emergence. This systematic variation is addressed by fitting a 2D polynomial to each of the components $[v_x, v_y]$ \citep[similar to][]{birch_2019} of the form $aX^2 + bXY + cY^2 + dX + eY + f$. This fitted polynomial, representing slowly varying (large-scale systematic) flows, is then subtracted from $[v_x, v_y]$ of every AR. The velocities for all the ARs are then averaged for a given $T_i$ and $div$ and $curl$ are obtained using equations~\ref{lctdiv} and~\ref{lctcurl}.

For the magnetograms, it is imperative to account for Joy's law and Hale's law (as described in section~\ref{section_introduction}). These are respectively taken care of by flipping the southern hemisphere magnetograms about latitude, and flipping the sign of the southern hemisphere magnetograms. Magnetograms for all the ARs are then averaged for a given $T_i$.

\section{Deep learning}
\begin{figure}[!htb]
    \includegraphics[scale=0.45]{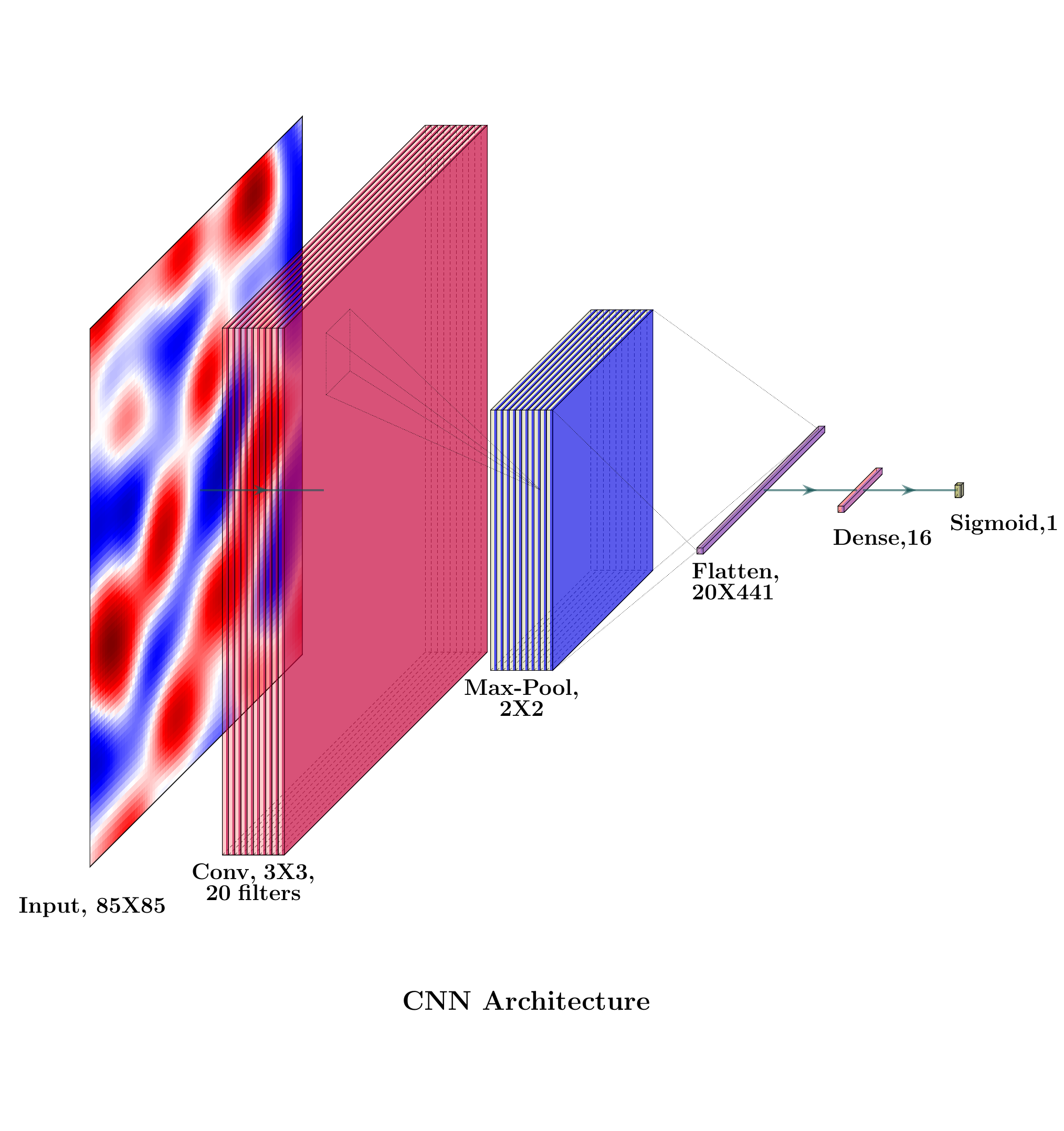}
    \caption{CNN architecture: From the left, representative input image (a horizontal divergence map), followed by Conv2D + Max-Pool2D, followed by Flatten, Dense, and the output Sigmoid layer. All the filters in the Conv2D layer operate independently on its input image. Filters of size 3X3 are used in the Conv2D layer - the 20 filters operate independently of each other, scanning the whole image, to produce 20 images of size 85X85 each.}
    \label{fig_NNarchitechture}
\end{figure}

Deep learning \citep{krizhevsky2012,goodfellow2016} is a set of machine-learning algorithms capable of locating hidden features and correlations in noisy data. In supervised learning, the machine is trained on labelled data to identify the implicit mapping between input-output pairs \citep{hastie_2001}. The convolutional neural network \citep[CNN,][]{lecun2015} is a class of deep-learning algorithms particularly adept at finding features in images by convolving them with multiple filters (convolution kernels). The architecture of CNN used in this work is shown in Figure~\ref{fig_NNarchitechture}. Various models, employing different configurations such as varying numbers of layers, activation functions, convolution kernel sizes, and learning rates, can statistically achieve identical results (within machine uncertainty of 2\% as reported in section~\ref{section_results}). Here we use one such architecture that is sufficiently optimal in terms of speed of training, depth of the neural network, their associated parameters, and computational demand .

It is natural to pose the problem of discerning between AR and QS (quiet sun) horizontal divergence images in the form of binary classification, with input-output pairs AR-1 and QS-0. Neural networks recognise patterns well when trained on abundant data sets. However, in the current setup, we only have 115 ARs from the observations. To expand our training dataset, we first collect a large number of images of QS horizontal divergence. We embed synthetic AR inflows (constructed from averaging over many supergranular inflows, as explained in the following Section~\ref{section_constructing_synthetics}) in some of these images (positives) and the rest are just the convective background (negatives), an approach that allows us to generate as many unique samples as needed for training. The entirety of the observed AR sample is preserved exclusively for testing. Our machine thus is designed to look for AR-like flow feature in images, and a failure to detect one will result in the machine associating an output 0 to that image. Ultimately, if there are features unique to active region emergence, it is expected that our formulation of the problem is equivalent to AR/QS classification.

In effect, once an image is passed through convolution filters, pooling and activation layers, the output neuron, termed \textit{sigmoid}, produces a number in the range $[0,1]$. The closer the output is to 0 or 1, the more confidently the machine indicates that the AR flow feature is absent/present. We classify all outputs $>0.5$ as containing the feature feature consistent with emergence (`positives'), whereas outputs $<0.5$ as `negatives'.

We train two independent deep-learning classifiers trained on horizontal divergence flow images - an ``inflow machine" to recognize pre-emergent AR inflows, and an ``outflow machine" to identify post-emergence AR outflows. A machine-learning model that is adequately trained to recognize AR-like flow features should  ideally mark all AR images as 1, and assign 0 to QS images. We conduct detailed analysis and interpretation of AR/QS flows in relation to magnetic-flux emergence based on the model outputs. Here, we primarily investigate horizontal divergent flows (see Figure~\ref{fig_main_ML}); machine learning results for radial vorticity and magnetograms are shown in Appendix ~\ref{supp_ML}. 

To summarize, our motive behind using machine learning in this work is to 
\begin{enumerate}
    \item appreciate how the flow and magnetic field features in ARs differ from QS areas,
    \item investigate if the flow features around the chosen 115 ARs are similar,
    \item explore the connections between different processes associated with flux emergence, i.e., examine if ARs that show strong bipolar fields also exhibit strong radial vorticity and/or horizontal divergence.
\end{enumerate}

We shall use the notation for the results of the machine on test data (see Table~\ref{TSS_synthetics}):
\begin{itemize}
    \item AR(1) - machine detects flow/magnetic field feature in an AR image,
    \item AR(0) - machine does not detect flow/magnetic field feature in an AR image,
    \item QS(0) - machine does not detect flow/magnetic field feature in a QS image,
    \item QS(1) - machine detects flow/magnetic field feature in a QS image.
\end{itemize}

\section{Generating Synthetics}\label{section_constructing_synthetics}
From previous statistical studies \citep{birch_2019,braun2016}, EAR horizontal divergence images presumably, on average, contain the weak, signature inflow close to the emergence location (see Figure~\ref{fig_main_fullset} B, first six panels). Based on this assumption, we will generate synthetic AR inflows using the algorithm described in \citet{birch_2019} (see for example Figure~\ref{fig_puresynthetics}, bottom row).

Using synthetics in training provides a great deal of flexibility in managing the training data. Benefits include the ability to generate a large, unique training sample with desired feature shapes at specified SNR. All the samples are first normalized by their absolute maximum value. Then we incorporate supergranular noise in synthetic AR inflows (see section~\ref{addnoise_subsection}), i.e., we add random QS images amplified by a factor chosen from $\mathcal{N}(6,1)$ in order to imitate the poor S/N of observed AR inflows ($\sim50$ m/s, compared to the $\sim$ 300 m/s background). The mean S/N of noisy AR flows here is thus $1/6$ (results for other values of S/N are shown in Figure~\ref{fig_ntp}). That is, during training, the machine is exposed to two categories of images only, synthetic AR flows embedded in realistic noise, and QS flows. Both of these datasets predominantly contain supergranluar flows.

We wish to highlight the other benefits derived by using synthetics in this work.
\begin{itemize}
    \item It allows for an unbiased analysis of the 115 observed ARs as these are not used in training the machine,
    \item Successful feature engineering implies we will have no shortage of training data - class imbalance issue \citep[e.g.][]{dhuri2020} is fully mitigated. That is, neural network training will not suffer due to unavailability of more than 115 observed ARs.
    \item As all the 115 ARs qualify as unseen data for the machine, we overcome the need to cross-validate (divide training data into subsets and train/test over different subsets to derive uncertainty). 
\end{itemize}

{AR flows evolve over time - Figure~\ref{fig_main_fullset} shows that EAR inflows are strongest / appear most prominently at -15 hrs., and rapidly change into outflows near and after emergence. Whereas the background convective field (outflow structures around the AR inflow) is a more slowly evolving phenomena, in line with the commonly reported lifetime of 1.5 days of supergranules \citep[see][]{rincon_rieutord_2018}. As the EAR inflow has been the flow structure previously associated with flux emergence, our main interest lies in replicating this pattern implanted in a supergranular convective flow field, with variance in the morphology, S/N, location, etc programmed for in the synthetics algorithm.}

{It is worth noting that the AR inflows for an ensemble of $\sim$110 instances are surrounded by the background convective field, which are stochastic realizations of supergranules not thoroughly cancelled upon averaging. Superimposing over a larger number of ``clean" emergence instances might well yield a more distinguishable flow feature, free from excessive background noise. Thus in this work, given what we observe in an ensemble averaged 110 ARs, we seek to build a machine that detects this most prominent flow feature in a slowly-varying, yet prominent background. The overall objective during training of the machine learning model is not to maximize the number of positive detections. It is rather to study the specific pattern that, in previous studies, has been correlated with magnetic flux emergence, and which can appear in all manner of convective flow-field. Subsequently from studying these AR flows in relation to the magnetic fields, we might hope to make statements about the driving forces behind flux emergence. Hence, we do not pay specific attention to the temporal evolution of stochastic background flows, but rather focus on generating realistic flow pattern of interest and embed it in quiet Sun fields (see section~\ref{addnoise_subsection}).}

\subsection{Flows}\label{syntheticfl_subsection}
Our goal in synthetics will be to emulate ensemble-averaged AR horizontal divergence images; procedure for radial vorticity remains the same.
\begin{itemize}
    \item Identify: features of supergranular scale of around 25-35 Mm using the \texttt{blob\_log} routine from the python package \texttt{skimage 0.19.2}. 
    \begin{itemize}
        \item Input: a flow map and amplitude threshold of the feature (which we set to be 1e-5 in an ad hoc fashion, as strong supergranular features in horizontal divergence flow images are roughly of this amplitude),
        \item Output: radius and center of the outflows/inflows.
    \end{itemize}
    \item Reject: from the list those features if, within their 120 Mm the smoothed ($\sigma$ = 5 pixels), unsigned magnetic field exceeds 120 G.
    \item Select: 20 features (randomly) from the remaining list.
    \item Repeat: for 50 QS images.
    \item Shift: features to desired region.
    \item Superimpose: all the 20*50=1000 features.
\end{itemize}
We then shift locations for the 1000 identified inflows are picked such that the final inflow is elongated in east-west direction with an offset to retrograde \citep{birch_2019}. A synthetic AR inflow is thus constructed by shifting and superimposing 1000 supergranular inflows, diminishing background supergranular noise by a factor of $\sim1/\sqrt{1000}$.

\subsection{Magnetic fields}\label{syntheticmf_subsection}
Generation of synthetic bipolar fields, similar to those seen in an ensemble-averaged AR magnetograms follows in a straightforward manner. Populate a region in a horizontal band ($\sim 30$Mm wide, chosen in an ad-hoc manner) around the center of a blank image with positive (negative) pixels in retrograde (prograde) direction. Ensure that the polarities close to each other, often times with slightly random tilts about the horizontal to account for Joy's law in observations.

\subsection{Adding background noise to synthetics}\label{addnoise_subsection}
\begin{figure}[!ht]
    \centering
    \includegraphics[scale=0.8]{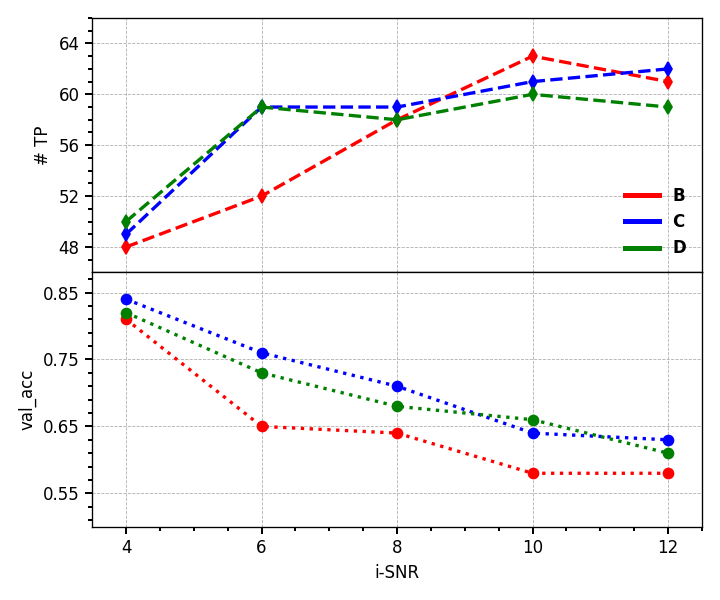}
    \caption{\textit{Top}: Number of positive predictions (TP) of the machine for bipolar fields B, radial vorticity C, and horizontal divergence D for the ARs in [-24, -18] hrs period. \textit{Bottom}: Accuracy on test data. Both these quantities are plotted as a function of different mean i-SNRs. Figure legend is common for both the panels.}
    \label{fig_ntp}
\end{figure}

The signal-to-noise of EAR inflows is poor (given that \citet{birch_2019} averaged 57 AR samples to image the pre-emergent inflows, SNR$\sim 1/\sqrt{57}$). We should also expect our machine to perform robustly on observations only if our synthetics possessed realistic background properties \citep[supergranular-scale flows and network fields,][]{dewijn2009}. These are satisfied by adding random QS images of same size and suitable amplitudes to pure features generated in \ref{syntheticfl_subsection} and \ref{syntheticmf_subsection} (see Figure~\ref{fig_synthetic}). We label the amplitude by which we scale this added QS image 'inverse-SNR' / 'i-SNR'. More the value of i-SNR, more the noise dominates the AR feature. We choose i-SNR from a random normal distribution $\mathcal{N}(\text{i-SNR}, 1)$ for the different AR images motivated by ensemble studies consideration, where it is only possible to quote the mean SNR of the feature.

i-SNR is a hyperparameter of the machine, in that we control it externally and it influences the prediction accuracy of the machine. Thus, it is natural to anticipate different training accuracy and predictions based on different input i-SNR. This is demonstrated for the ensemble-averaged EAR images of bipolar fields, radial vorticity and horizontal divergence, in the [-24, -18] hrs. period in Figure~\ref{fig_ntp}. We interpret this figure as follows: the machine learns to detect increasing numbers of poor-SNR AR features as it is trained on noisier synthetic AR images (top panel showing increasing number of positive ("TP"/ True Positive) detections with increasing i-SNR/noisier data), but at the same time, it becomes more susceptible to false-positive predictions in QS images (suggested by the bottom panel, with the drop in validation accuracy with increasing i-SNR/noisier data).

\section{Training and testing on synthetics}
Although the goal is to understand the predictions of machine on observed AR images, it is useful to first illustrate the predictions on synthetic ARs and observed QS.
\subsection{Training}
Datasets in machine-learning are conventionally split into ``training", ``validation", and ``test". During training, the machine is only allowed access to training data and does not use validation data to adjust the weights and biases of the neural network. Rather, at every epoch, machine performance is evaluated on validation data. Therefore, the criteria that all the samples in the test dataset remain unseen by the machine during training may be satisfied by using the same dataset for ``validation" and ``test". Below are the train/test data split and model parameters.

\begin{itemize}
    \item Train: 10000 images - 5000 AR, 5000 QS,
    \item Test: 2000 images - 1000 AR, 1000 QS.
    \item learning rate: $10^{-5}$,  epochs: 30,  batch\_size: 30,
    \item optimizer: Adam, loss: \texttt{binary\_crossentropy}, metric:\texttt{binary\_accuracy}.
\end{itemize}

% \begin{figure}[!htb]
%     \includegraphics[scale=0.65]{acc_loss.png}
%     \caption{Model accuracy and loss with increasing number of epochs, for both training and test data (containing synthetic ARs and QS).}
%     \label{fig_acc_loss}
% \end{figure}

\subsection{Testing}
\begin{figure}[!htb]
    \centering
    \includegraphics[scale=0.55]{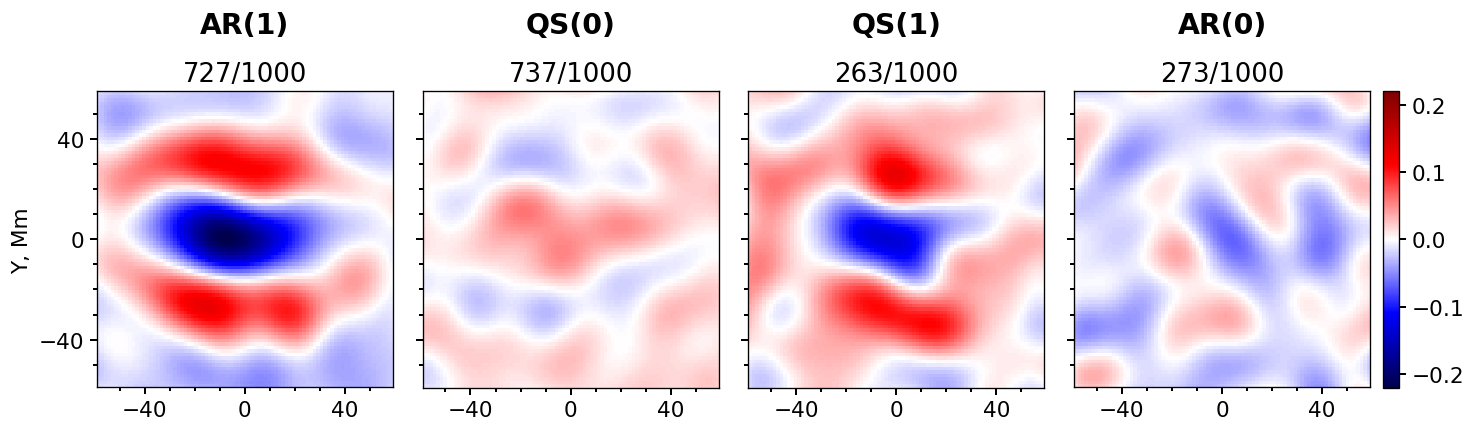}
    \caption{Average of the images belonging to the four outcomes. Figures in a row are saturated to the colorscale of AR(1) (the first image in a row). The above results are summarized in Table~\ref{TSS_synthetics}. Results corresponding to magnetic bipolar fields and radial vorticity are show in Appendix Figure~\ref{fig_class_syn_BC}.}
    \label{fig_class_syn_D}
\end{figure}

The machine is tested on 1000 synthetic AR and 1000 observed QS images. Table~\ref{TSS_synthetics} summarizes the results for test dataset, with i-SNR chosen from the random normal distribution $\mathcal{N}(6,1)$. AR(1) and QS(0) in Figures~\ref{fig_class_syn_D} (and~\ref{fig_class_syn_BC}) (the first two columns in the figure) are where features are correctly detected to be present and absent, respectively. Two other results that the test on synthetic AR and observed QS demonstrate: 1) As seen from the average of all the AR(1) and QS(1) in Figure~\ref{fig_class_syn_D} (and~\ref{fig_class_syn_BC}), the model has learned to identify the desired magnetic and flow features in both active region and quiet Sun images. 2) It also misses detection when the feature is dominated by large background noise, as evidenced by the average of all the AR(0). As explained in ~\ref{addnoise_subsection} and Figure~\ref{fig_ntp}, increasing i-SNR might increase AR(1) in observed AR but will simultaneously increase QS(1) in QS, reducing overall accuracy. The discussion on accuracy is limited to testing on synthetics merely as an illustration, as our primary aim is to study if the AR-like flow and magnetic field features are unique to active regions alone.

True Skill Statistics (\textit{TSS}) provides a measure of the accuracy of machine in feature detection / classification tasks. It is computed using the rate of correct detections of both the classes (1 \& 0) in the test sample, that is, how many AR and QS are mapped to 1 and 0 as a fraction of their total sample size, respectively. For our case, we use the below formula.
$$TSS = \frac{AR(1)}{AR(1) + AR(0)} + \frac{QS(0)}{QS(0) + QS(1)} -1$$

\begin{table}[h!]
\setlength{\arrayrulewidth}{0.4mm}
\setlength{\tabcolsep}{15pt}
\renewcommand{\arraystretch}{1.2}
\centering
\begin{tabular}{c||ccccc}
\hline
\hline
feature & AR(1) & QS(0) & QS(1) & AR(0) & \textit{TSS} \\
\hline
horizontal divergence & 728 & 728 & 272 & 272 & 0.456 \\
\hline
bipolar fields & 669 & 755 & 245 & 331 & 0.424\\
\hline
radial vorticity & 720 & 729 & 271 & 280 & 0.449\\
\hline
\end{tabular}
\caption{Results for synthetics. Figures corresponding to the above is shown in Figure~\ref{fig_class_syn_D} (and~\ref{fig_class_syn_BC}). Given in the last column are accuracy scores, also known as True Skill Statistics (TSS)}
\label{TSS_synthetics}
\end{table}

\section{Results on observations and interpretation}\label{section_results}
\begin{figure}%[!htb]
    \includegraphics[scale=0.5]{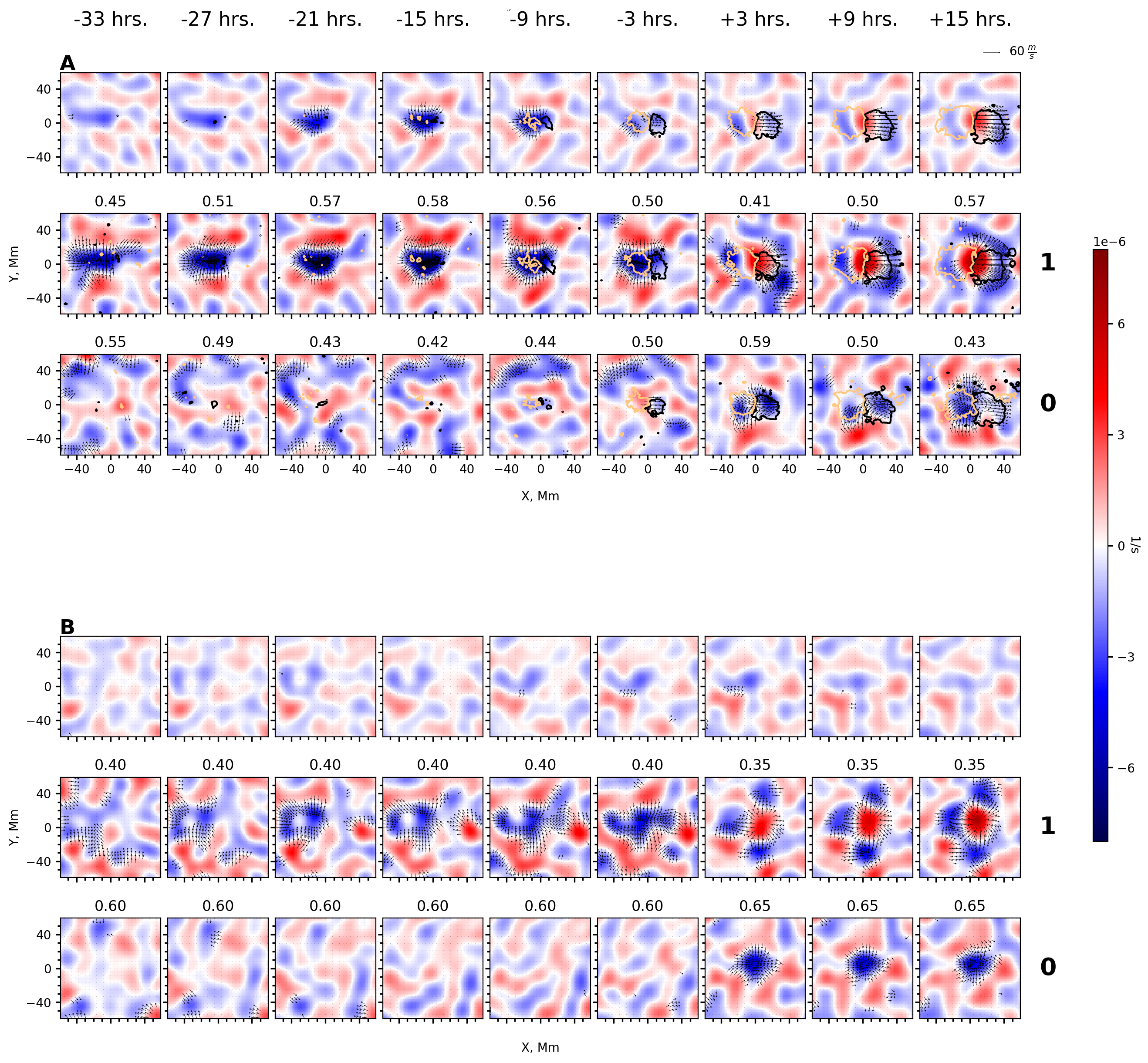}
    \caption {Machine learning classification of horizontal divergence flow images into 1/0, or as images with / without AR-like flow features. Panels \textbf{A}, \textit{top row}: ensemble averages over all ARs in a given interval, \textit{middle row}: averages over ARs where flow features are detected, and \textit{bottom row}: averages over ARs where flow features are not detected by the model. Magnetic field contours of $\pm 10$ G are overlaid. Panels \textbf{B}: Same as \textbf{A}, except all the images are QS (110 samples). The fraction of total samples classified as 1/0 are stated at the tops of the middle and bottom rows in both \textbf{A} and \textbf{B}.}
    \label{fig_main_ML}
\end{figure}

To make predictions in the observations, we use the inflow machine at pre-emergence times ($t<0$), and the outflow machine on post-emergence ($t>0$) flow images. The model outputs for the ARs in the nine time intervals are noted. AR flow images in each interval are categorized containing or not showing flow signal (machine output 1/0). The results are plotted in Figure~\ref{fig_main_ML}A.

Contrary to the impression gained from average measurements in ensemble studies, only some ARs contain inflow signatures. We find that the category of `AR flows' is broad, i.e., there exists a sub-class of ARs that do not show any flow features. In each time interval, the contrast between the middle and bottom rows in Figure~\ref{fig_main_ML}A shows that the model was able to pick out those particular ARs that contain the AR-like inflow ($t<0$) and outflow ($t>0$) feature from the entire set. The fraction of total ARs in a given interval with or without these AR-like flow features is shown at the top of each panel in the middle and bottom rows. Based on this model, anywhere between 40-60\% of ARs contain the feature. For instance, in the fourth column ($T_i = -15$ hrs.), there are 107 ARs, of which 58\% (62 ARs) contain the inflow, whereas it is absent in the other 42\% (45 ARs). We trained multiple independent machines, on training samples generated afresh each time, to check for consistency in the predictions, i.e., to test whether the same set of ARs is categorized as 1/0. We find that the predictions remain consistent to within 3\%, i.e., for an observational sample size of 100, outputs for only 3 ARs vacillate around 0.5 in the sigmoid output, where in different models, one or more of these 3 may switch between 1/0 - samples that are close to the decision boundary since they are harder to classify. This is understood to be machine error, rather than the actual noise statistics of observed AR flows.

A cursory glance at the middle row of Figure~\ref{fig_main_ML}A reveals that ARs with pre-emergence inflows seemingly evolve into outflows post-emergence. To verify if this is the case, we plot a Venn diagram (Figure~\ref{fig_venn_statistics}A) for two representative sets - (62) ARs that show inflows at -15 hrs. and (64) ARs that show outflows at +15 hrs. We refer to these two sets as `inflow ARs' and `outflow ARs', respectively. Roughly equal numbers of ARs are present in the three regions of the Venn diagram, i.e., ARs with pre-emergence inflows may or may not show outflows post-emergence. 

The overlaid magnetic field contours in the middle and bottom rows of Figure~\ref{fig_main_ML}A indicate an absence of meaningful correlation between flow features and magnetic fields, i.e., the presence of bipoles has little-to-no bearing on the presence of inflows/outflows in their vicinity.

\subsection{Weak evidence for the existence of signature flows in ARs}
In order to ascribe flow features as ARs, a comparison with QS flows allows for establishing a baseline, the threshold above which flow features may be correlated to large-scale magnetic fields with appropriate confidence. For this purpose, we collect 110 QS horizontal divergence flow images, evolve them for 54 hours / nine contiguous 6-hr intervals (the same duration over which flows around ARs are studied). To place error-bars, we evolve 30 different batches of 110 samples over the nine intervals to obtain the mean and standard deviations.

Figure~\ref{fig_main_ML}B shows the ensemble average of 110 samples (top row) and the fraction of samples with/without identified AR-like flow features in the middle and bottom rows, respectively. The CNN model does not recognize AR-like flow features as unique to only active regions. There is a non-negligible baseline rate (35-40\%) with which these features also appear in quiet-Sun flow fields. Since we train our model on synthetic flows, which are averaged over many supergranules in QS, it may be predisposed to finding AR-like flow features in QS. That the CNN model finds these features in both AR and QS images indicates that flows associated with active regions may not be physically distinct from background supergranular flows. 

\begin{figure}%[!htb]
    \includegraphics[scale=0.62]{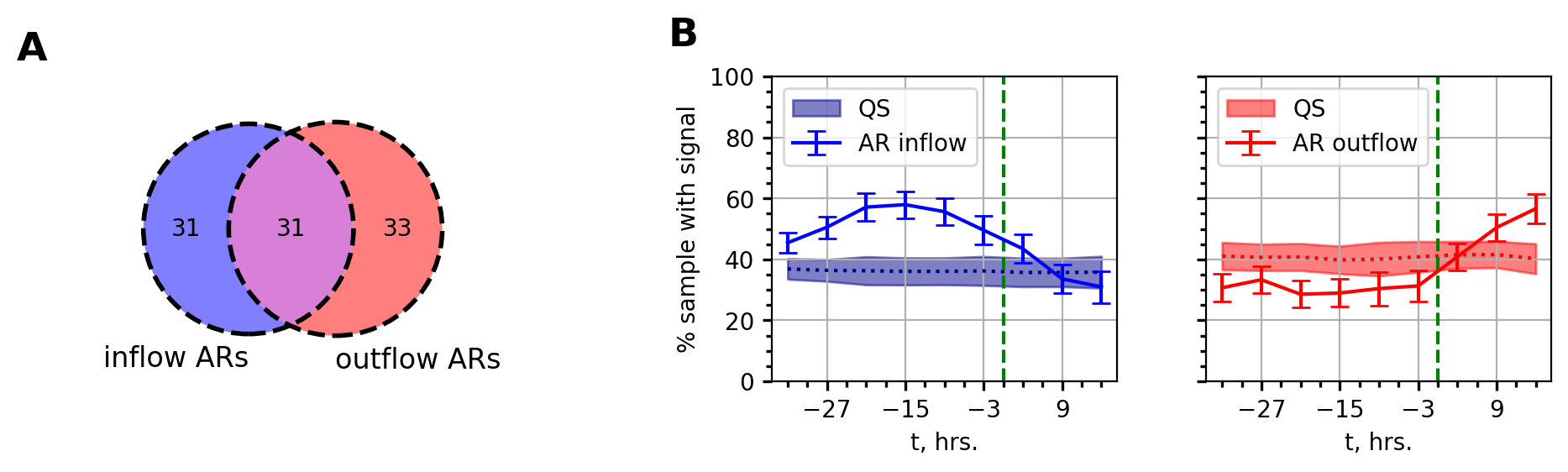}
    \caption{\textbf{A}: Venn diagram illustrating the correlation between ``inflow ARs" and ``outflow ARs". \textbf{B}: Statistics of AR (solid lines) and QS (shaded region) vs time, showing the fraction of total samples containing AR-like inflows (left panel) and AR-like outflows (right panel). The vertical green dashed line marks the emergence time $t=0$.}
    \label{fig_venn_statistics}
\end{figure}

In Figure~\ref{fig_venn_statistics}B we compare the rate with which AR-like inflows and outflows (fraction of total samples in which flow features are detected by the CNN model) appear in both AR and QS vs time t before and after emergence. For sake of completeness, ARs in all nine intervals are passed through both inflow and outflow CNN models to estimate the corresponding occurrence rates over the course of emergence. 30 batches of 110 QS samples are also evolved and tested by the CNN model to obtain mean and 1-$\sigma$ standard deviations of the quiet-Sun rates (shaded blue and red curves on the left and right panels). The AR prediction rate is almost constant with time for the QS maps, indicating that the background is statistically time-invariant. We apply the same QS error-bars to AR rates, with the assumption that noise statistics are similar. Consequently, these features are present in only a fraction of AR samples, and the statistical significance of the rate with which they appear above the background ($\simeq3$-$\sigma$ only at -15 hrs.) may be debatable, indicating the lack of robust emergence-related flow signatures (from the current sample size). The results indicate a weak tendency for flux to emerge near supergranular inflows. Surface AR-inflow amplitudes of 60 m/s are much weaker than supergranular speeds of $\sim300$ m/s, pointing towards an imperfect alignment of emergence location and supergranular cell boundaries. While the correlation between AR inflows and emergence location and supergranular boundaries was explored in \citet{birch2013} and \citet{birch_2019}, conclusive evidence was lacking until present.

\subsection{Weak evidence for magnetic buoyancy-assisted emergence}
Next, we carry out tests to explore two phenomena predicted to be associated with magnetic buoyancy-assisted flux emergence: (1)  a hypothesised retrograde flow \citep{fan2008,weber2011} ($-v_x$ feature) that appears in numerical simulations and has been explained as arising due to angular momentum conservation, and  (2) an enhanced time rate of change for flux (appearing at the surface due to additional upward force), as compared with the purely convectively driven flux scenario. In numerical simulations of buoyancy-boosted flux emergence in near-surface layers \citep{cheung2010,rempel2014}, and in observations \citep{toriumi2012}, outflows of the order of km/s magnitude \citep[as compared to inflows of $\sim 50$ m/s amplitudes seen in][]{birch_2019} are seen near the emergence location. Therefore, we check if outflow ARs show retrograde flows and/or elevated flux growth rates at the surface compared to inflow ARs, a possible sign that magnetic buoyancy is a driver.

\begin{figure}%[!htb]
    \includegraphics[scale=0.52]{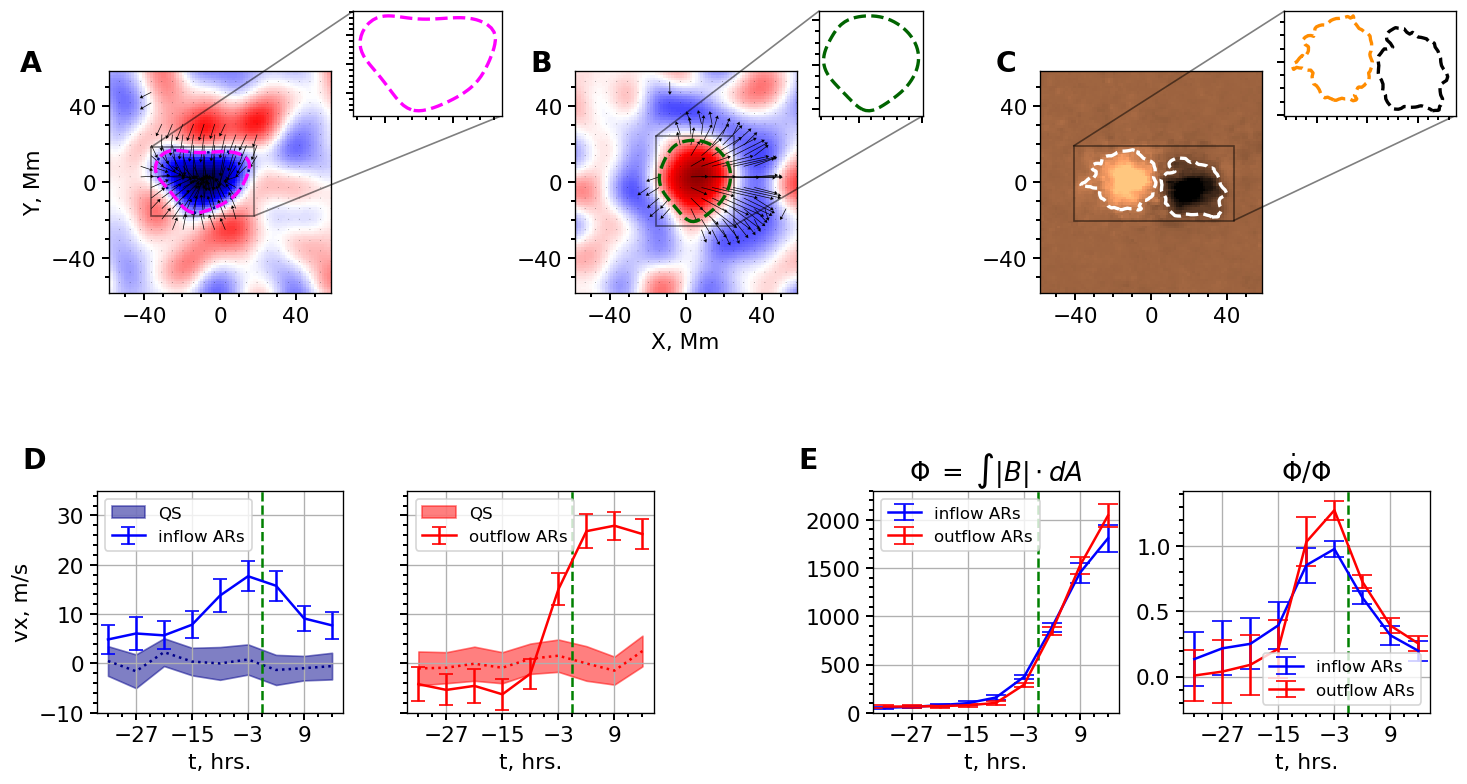}
    \caption{\textbf{A}: contour for $\overline{v_x}$ for inflow ARs. \textbf{B}: contour for $\overline{v_x}$ for outflow ARs. \textbf{C}: contour for flux and flux rate of change. \textbf{D}: illustration of $\overline{v_x}$ near emergence location (solid lines) in ``inflow ARs" (left panel) and ``outflow ARs" (right panel) vs time, along with $\overline{v_x}$ in QS (shaded region) inside an area of the same contour as chosen for the ARs. \textbf{E}: Unsigned flux $\Phi$ (left panel), and time rate of change of $\Phi$ (right panel), for ``inflow ARs" and ``outflow ARs". The vertical green-dashed line marks the emergence time $t=0$. Error-bar heights in \textbf{D} and \textbf{E} are $\pm 1 \sigma$.}
    \label{fig_contour_buoyancy}
\end{figure}

The average velocity in the $x$ direction, $\overline{v_x}=\displaystyle\frac{\int v_x \;\textrm{d}A}{\int \textrm{d}A}$, is computed as the average over a contour enclosing the inflow region for the 62 inflow ARs, and outflow region for the 64 outflow ARs (Figure~\ref{fig_contour_buoyancy}A and B), as these are the characteristic flow features of these two sets of ARs. We only include a sufficiently strong boundary around the inflow / outflow; weak portions of the inflow / outflow feature merge into the neighbouring convective background features and we therefore choose to avoid including them. This is achieved by trial and error and setting the threshold to be $1.5\times10^{-6}$s$^{-1}$ in the horizontal divergence image.

We obtain $\overline{v_x}$ vs time for both inflow and outflow ARs by evolving them over the nine intervals and plot them in Figure~\ref{fig_contour_buoyancy}D as solid lines. To ensure that the $\overline{v_x}$ in ARs is above the background, we compute QS values of $\overline{v_x}$. We gather 30 batches of 62 QS images containing AR-like inflows (output 1 in the inflow machine), and 30 batches of 64 QS images containing AR-like outflows (output 1 in outflow machine) over the nine intervals. Mean and 1-$\sigma$ standard deviations of the baseline $\overline{v_x}$ are plotted as shaded regions in the two panels. We again place the same QS error-bars on the AR $\overline{v_x}$. We measure no statistically significant retrograde flow (negative $\overline{v_x}$) at the surface in either outflow or inflow ARs. However, we detect a strong prograde flow in outflow ARs, post-emergence, that might be correlated with the leading polarity of the bipoles moving faster away from the polarity inversion line than the following polarity \citep{schunker_2019}.

To check if flux emerges at a more rapid rate in outflow ARs compared to inflow ARs, we choose a contour such that it sufficiently encloses the bipoles in the ensemble averaged magnetograms of these two sets (Figure~\ref{fig_contour_buoyancy}C). We then compute $\Phi=\int |\bf{B}|\;\textrm{d}A$ within that contour, using the same contour for both the sets, which is in turn obtained from the $\pm25$ G contours associated with the average of 113 magnetograms at +15 hrs. $\Phi$ is plotted in Figure~\ref{fig_contour_buoyancy}E, left panel. To place 1-$\sigma$ error-bars, we estimate the standard deviation from within the 62 inflow-AR magnetograms and 64 outflow-AR magnetograms. Next, to compare the time rate of change of flux, $\displaystyle\frac{d\textrm{ln}\Phi}{dt}$ is obtained (right panel), which may tell us if the flux of outflow ARs rises faster, i.e., if their net unsigned flux at the surface increases faster than that of inflow ARs before emergence time. 
Figure~\ref{fig_contour_buoyancy}E shows the time rate of change of flux for the small number of outflow and inflow ARs in our data set. We find that for most time periods the rate of change is consistent between the inflows and outflows. However, at $T=-3$ hrs. the outflow's rate of change is greater than the inflow by 2-3~$\sigma$. This coincides with the maximal time rate of change for both inflow and outflow ARs. We note that the small number of samples available for both inflow and outflows likely leads to poor estimates of the error bars, and caution must be taken when interpreting the significance of these two data points. Overall, the time rate of change of flux in the emerging inflow and outflow ARs is fairly consistent (compared to the order of magnitude difference suggested by previous authors) lending little support to the hypothesis that outflow ARs, enhanced by magnetic buoyancy, emerge considerably faster.

\section{Discussion: Comparison with simulations}
Helioseismic studies of EARs, which attempt to infer pre-emergent signatures below the surface using acoustic waves, suffer from low S/N beneath the photosphere, leading to a large spread among the findings (see for instance the introduction section of \citet{birch2013}). It is not possible to make assertions about the subsurface flow and magnetic field dynamics from this present study, although we may possibly rule out flux emergence models that misalign with our observations. A major difficulty in modeling the flux emergence process over the vertical extent of the convection zone is the stark density contrast between the top and the bottom layers, placing a steep cost on numerical calculations. Simulations are thus carried out sometimes by only modeling emergence in the top $\sim$ 20 Mm (including the photosphere), over which density drops sharply and the thermodynamics of the plasma is complicated by ionization and radiative effects \citep{nordlund_etal_2009}. A challenge in these setups is the initiation of flux emergence, since these simulations do not capture large-scale dynamo processes that produce the emerging magnetic field in the first place. Common approaches are passive flux emergence by imposing magnetic field in convective inflow regions, active flux emergence in which a flux structure (typically a semi-torus) is driven into the simulation domain across the bottom boundary, and the insertion of a buoyant flux tube. The details of the resulting flux emergence in terms of speed are dependent on the specifics of the setup. We briefly review results from relevant, near-surface simulations to compare with our inferences.

Radiative MHD simulations of uniform, untwisted, weak (1 kG) magnetic field rising from 20 Mm below the photosphere \citep{stein2011,stein2012} (passive flux emergence) find the rise speed to be of the same order as the simulated convective upflows at these depths. Their study further implies that downflow lanes in intermediate-sized convective cells (presumably supergranular boundaries) mainly serve to dictate the locations where bipolar fields form, commensurate with the findings of simulations of AR formation in global-scale solar convective dynamos \citep{chen2017}. Our conclusions also moderately align with this scenario. Supergranular boundaries and emergence locations are correlated (although weakly). Moreover, \citet{stein2012} suggest that the combined actions of up and downflows on flux tubes in a supergranular-sized region ($\sim$ 30 Mm) is sufficient for them to form ARs of similar sizes, i.e., magnetoconvection alone is enough to produce ARs. 

Active flux emergence setups in which a magnetic half-torus was kinematically driven across the bottom boundary condition were considered by \citep{cheung2010,rempel2014}. \citet{birch2016} showed that the resulting horizontal flows were too strong compared to observations unless vertical flows associated with flux emergence were comparable to typical convective upflows ($<150$ m/s in a depth of 20Mm), also indicative of a more passive flux emergence process.

Recent attempts \citep{toriumi2019_2,hotta2020} to overcome the drawbacks of flux-emergence simulations over limited vertical extents have focused on expanding the domain to cover the entire convection zone and to allow the dynamics of the flux tube to naturally play out through the interaction of convective flows and magnetic fields. Unlike previous setups, they started from a buoyant magnetic flux tube inserted in the volume of the simulation domain. They found that the rise speed of the flux tube is $\sim250$ m/s at 18 Mm below the photosphere; this exceeds the upper limit of 150 m/s at 20 Mm reached in a study \citet{birch2016} of magnetoconvection simulations constrained by observed surface flows around 70 active regions. Recent follow-up work by \citet{kaneko2022} showed that flux emergence is strongly influenced by the interaction with convective flows throughout the convection zone. To that end, they repeated their flux emergence setup more than 90 times, exploring a variety of initial locations for the flux tube.

The lack of correlation between AR-like flow features and large-scale magnetic fields, as seen from the present analysis, suggests that flux emergence is dominated neither by supergranular-scale convective flows nor by magnetic buoyancy. Other near-surface simulations \citep{cheung2010,rempel2014,chen2017}, reasonably in line with conclusions of previous \citep{stein2011,stein2012} studies, have found that supergranular-scale mean flows tend to oppose the assimilation of magnetic elements into similar polarities. This assimilation is counteracted by a Lorentz force due to correlation between small-scale fluctuations in velocity and magnetic fields. We thus speculate that flux emergence and the formation of coherent bipoles and monolithic sunspot structures are driven by small-scale turbulent flows (of the order of few granules, or a few Mm length scale), rather than large-scale mean flows with supergranular length scales ($\sim30$Mm), as obtained here using LCT. Although here we only analyze near-surface flows, convection in the deeper layers (below 20 Mm), imaged using seismic techniques, may also likely play a role in setting up the observed active-region scale magnetic fields at the surface.

\section*{Acknowledgements}
The machine-learning part of the research was carried out in the Intel® Xeon® CPU E5-2620 v3 GPU in the Department of Astronomy and Astrophysics at Tata Institute of Fundamental Research, Mumbai, India. Parallel processing for analysis of intensity and magnetogram files were done on the Intel® Xeon® Platinum 8280 CPU. 

\textbf{Funding:} This material is based upon work supported by the National Center for Atmospheric Research, which is a major facility sponsored by the National Science Foundation under Cooperative Agreement No. 1852977.
This research was supported in part by a generous donation (from the Murty Trust) aimed at enabling advances in astrophysics through the use of machine learning. Murty Trust, an initiative of the Murty Foundation, is a not-for-profit organisation dedicated to the preservation and celebration of culture, science, and knowledge systems born out of India. The Murty Trust is headed by Mrs. Sudha Murty and Mr. Rohan Murty.
This material is based upon work supported by Tamkeen under the NYU Abu Dhabi Research Institute grants G1502 and CASS.

\textbf{Author contributions}: S.M.H , P.M, and C.S.H designed the research. P.M. performed all the analysis. C.S.H. obtained data. S.D. and P.M. interpreted the machine learning results. M.R. helped with the conclusion. 

\textbf{Data and materials availability:} The HMI data are available on the JSOC export site {http://jsoc.stanford.edu}. The codes for machine-learning, generating synthetic flows and magnetic fields are available from the author upon request.

\textbf{Software}: pyFLCT \citep{welsch_etal_2004,fisher_welsch_2008,sunpy_community2020}, tensorflow \citep{tensorflow2015-whitepaper}, numpy \citep{harris2020array}, scipy \citep{2020SciPy-NMeth}, skimage feature dection \citep{van2014scikit}, netdrms v9.3\footnote{http://jsoc.stanford.edu/jsocwiki/DRMSSetup}, mtrack v2.6 \footnote{https://hmi.stanford.edu/rings/modules/mtrack.html}

\appendix

\section{pyFLCT}\label{section_pyflct}
\texttt{pyFLCT} takes as input a pair of intensity images with time separation (45 s) between adjacent images. The routine is designed to focus on localized regions in the image pairs by fading out the contribution from areas outside of the subregion and obtain velocities for such sub-images. This process is repeated until the whole image is covered. The extent of localization desired is given by the input \texttt{sigma}, which we set to 5 pixels ($\sim2.5$ Mm). We run the routine over a given time interval $T_i$, where $i={1...9}$, and obtain average velocity for that interval. As we are only interested in supergranular scale features, we smooth $[v_x, v_y]$ with a Gaussian of full-width 10 Mm and then obtain horizontal divergence and radial vorticity (see equations ~\ref{lctdiv} and~\ref{lctcurl}).

\section{Input image size}\label{input_image_subsection}
\begin{figure}[!htb]
\centering
    \includegraphics[scale=0.7]{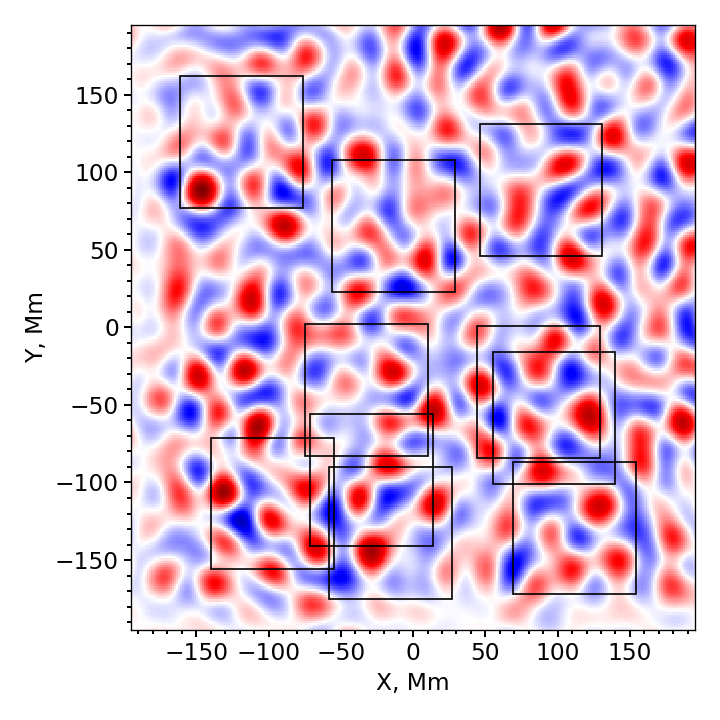}
    \caption{A representative $32^\circ\times32^\circ$ QS horizontal divergence flow map. Black squares overlaid on the map indicate random $10^\circ\times10^\circ$ oversampling. While there is of course a chance of partial overlap between smaller images, machines see them as unique samples. This is also done for bipolar fields and radial vorticity images.}
    \label{fig_oversample}
\end{figure}

Our tracked data products of 115 ARs are $32^\circ\times32^\circ$ in spatial size. The region of interest is only near the center of the image where active region is set to emerge; prior studies \citep{birch_2019, gottschling_2021} have shown flows associated with EARs to be limited ($\sim 7^\circ$) in spatial extent. Therefore we use images of a smaller size, of $10^\circ\times10^\circ$, for training the machine. Moreover, we interpolate the images onto a coarser grid of resolution 1.4 Mm, as small scale features are not relevant for successful training. Synthetic AR images are generated on this grid and QS images are randomly oversampled from the larger image to reduce computation load. That is, instead of obtaining multiple separate $10^\circ\times10^\circ$ QS observations, 150 different images of the same area are obtained from a $32^\circ\times32^\circ$ image in a process called oversampling (see Figure~\ref{fig_oversample}).

\begin{figure}
    \centering
    \includegraphics[scale=0.7]{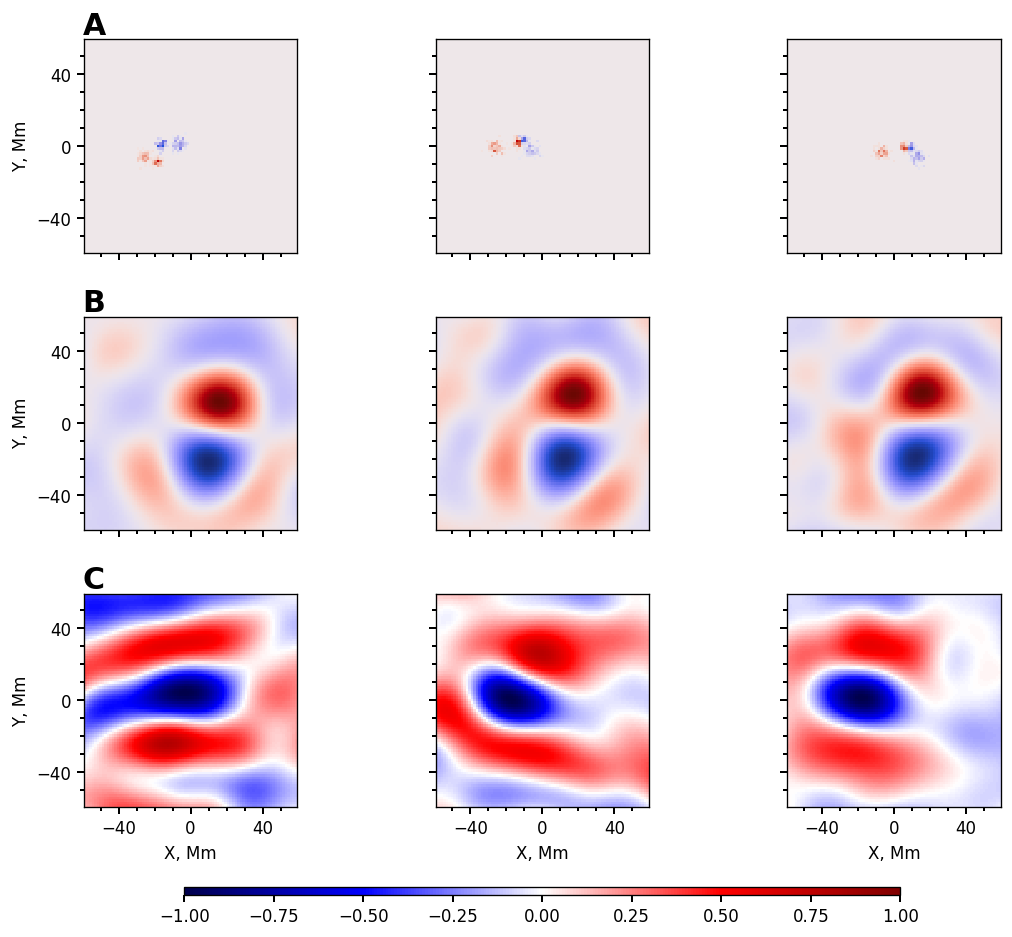}
    \caption{Examples of synthetic bipolar fields images (\textit{top row}), synthetic radial vorticity images (\textit{middle row}), and synthetic horizontal divergence images (\textit{bottom row}).}
    \label{fig_puresynthetics}
\end{figure}

\begin{figure}[!htb]
    \centering
    \includegraphics[scale=0.65]{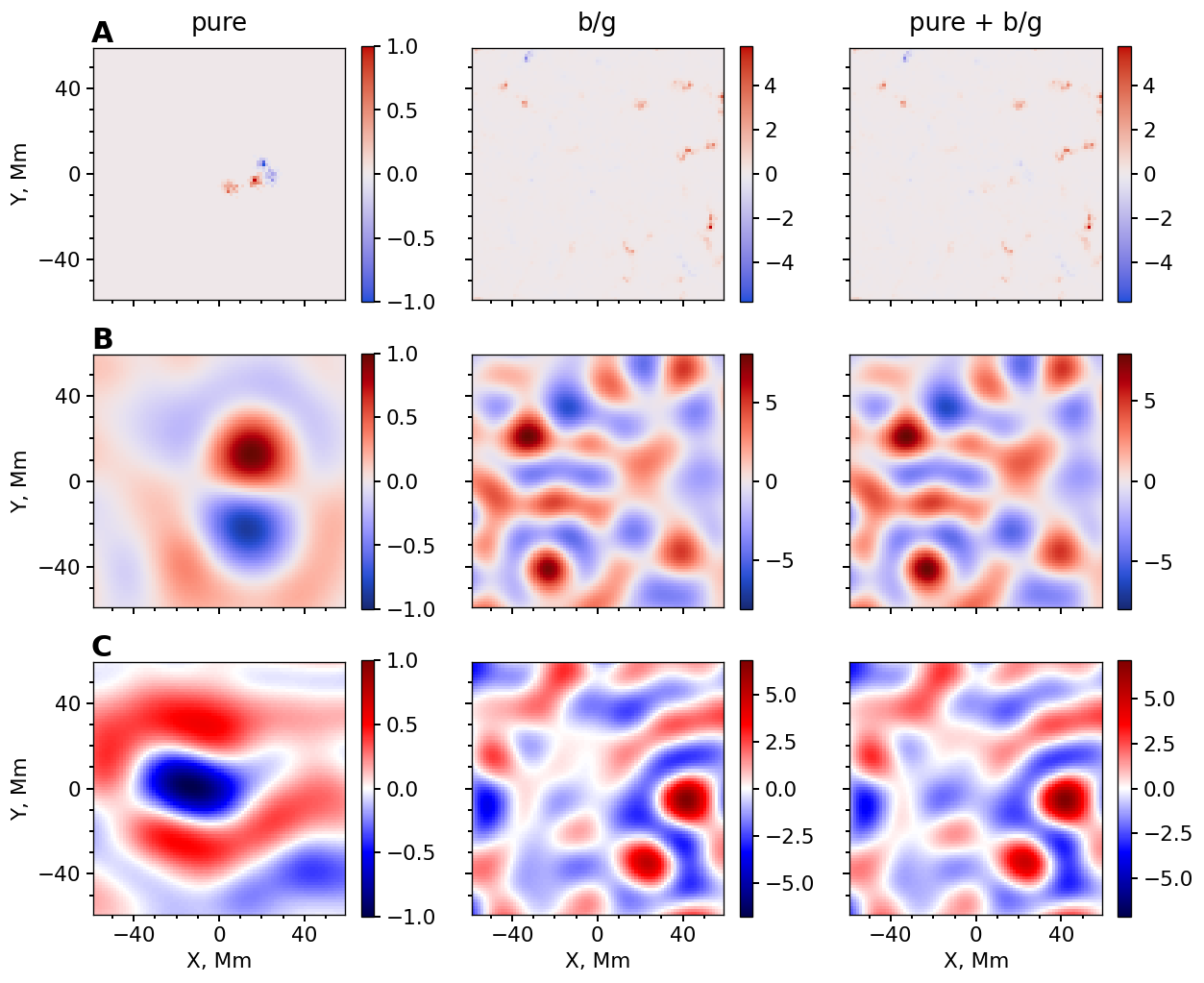}
    \caption{\textit{Top row}: Magnetic field, \textit{Middle row} vorticity, \textit{Bottom row}: inflows. \textit{Left column}: Pure features generated using procedure in \ref{section_constructing_synthetics}, \textit{Middle column}: Observed QS (background), with amplitudes greater than pure features, to imitate signal-to-noise in observations, \textit{Right column}: Adding left and middle column images and normalizing to 1.}
    \label{fig_synthetic}
\end{figure}

\section{Complete test results on synthetics and observation}\label{supp_ML}
\begin{figure}%[!htb]
    \includegraphics[scale=0.5]{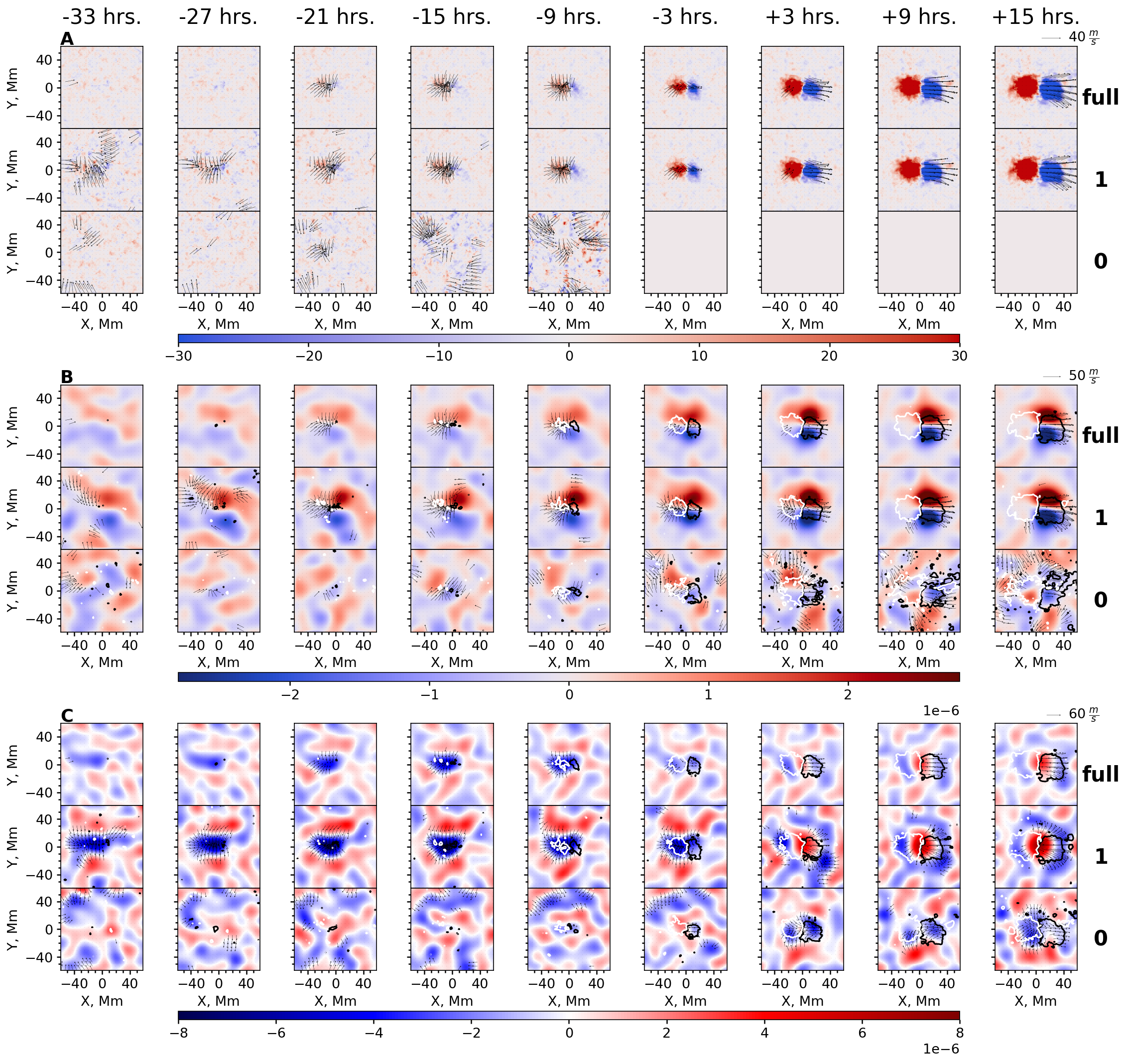}
    \caption{Machine-learning results for observations: \textbf{A} Line-of-sight magnetic field (in units of Gauss), \textbf{B} Radial vorticity (in units of 1/s), and \textbf{C} Horizontal divergence (in units of (1/s)). In each panel, full, 1, 0 denote ensemble averaging of all the AR, ARs in which machine detected the flow/magnetic field feature, ARs in which it did not detect flow/magnetic field feature.}
    \label{fig_obs}
\end{figure}

The trained model is tested on bipolar fields, radial vorticity, and horizontal divergence images, in the nine time intervals $T_1$ to $T_9$, for all the ARs in a given interval. We do not show results on QS here again. The number of available ARs decreases the farther back from emergence time one observes, as ARs tend to move outside of the desired field of view ($\pm 50^\circ$ central meridian distance). The total number of ARs for the nine $T_i$'s, along with the number of 1 and 0 detections are tabulated for bipolar fields, radial vorticity, and horizontal divergence images, are noted down in Table~\ref{table_classified_observation}. Figures corresponding to Table~\ref{table_classified_observation} are shown in Fig~\ref{fig_obs}. We draw the F
following conclusions from the results -
\begin{enumerate}
    \item The model shows all 115 ARs as containing bipolar magnetic fields from $T_6$, which is the interval closest to emergence time and forward (see Figure~\ref{fig_obs}A, last three columns). 
    \item Number of ARs predicted by the machine as containing bipoles (bipolar fields) and double vortex rolls (radial vorticity) steadily increases with time. But the number of ARs containing inflows/outflows (horizontal divergence) peaks at -15 hrs. and +15 hrs., respectively.
    \item Strength of magnetic field and vorticity, in bipolar fields and radial vorticity images respectively, also steadily increases with time, while the strengths of the inflows in horizontal divergence images peak around -15 hrs. 
\end{enumerate}

\begin{figure}[!htb]
    \centering
    \includegraphics[scale=0.55]{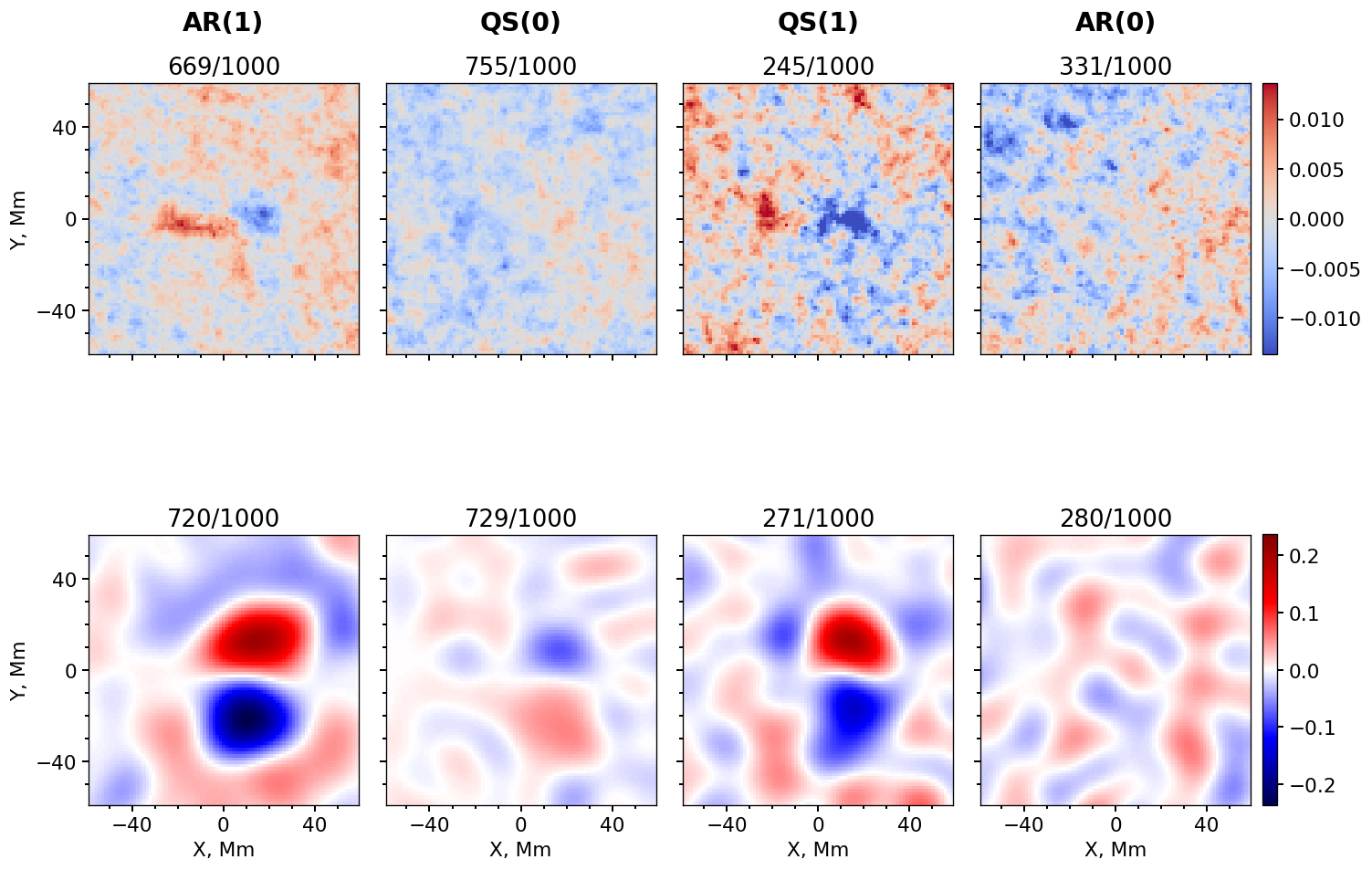}
    \caption{Machine-learning results for synthetics: Average of the images belonging to the four outcomes. \textit{Top row}: Magnetic bipolar fields. \textit{Bottom row}: Radial vorticity. Figures in a row are saturated to the colorscale of AR(1) (the first image in a row). The above results are summarized in Table~\ref{TSS_synthetics}.}
    \label{fig_class_syn_BC}
\end{figure}

\begin{table}[h!]
\setlength{\arrayrulewidth}{0.4mm}
\setlength{\tabcolsep}{8pt}
\renewcommand{\arraystretch}{1}
\centering
\begin{tabular}{c||c|c|c|c|c|c|c|c|c}
\hline
\hline
 & -33 hrs.  & -27 hrs.  & -21 hrs.  & -15 hrs.  & -9 hrs.  & -3 hrs.  & +3 hrs.  & +9 hrs. & +15 hrs.  \\
total & 88 & 93 & 98 & 107 & 115 & 115 & 115 & 113 & 113\\
\hline
bipolar fields & 1: 32 & 1: 42 & 1: 54  & 1: 80 & 1: 103 & 1: 115 & 1: 115 & 1: 113 & 1: 113 \\
      & 0: 56 & 0: 51 & 0: 44  & 0: 27 & 0: 12  & 0: 0  & 0: 0  & 0: 0  & 0: 0\\
\hline
radial vorticity & 1: 43 & 1: 45 & 1: 52  & 1: 60 & 1: 68 & 1: 86 & 1: 93 & 1: 94 & 1: 91 \\
      & 0: 39 & 0: 48 & 0: 46  & 0: 47 & 0: 47 & 0: 29 & 0: 22 & 0: 19 & 0: 22\\
\hline
horizontal divergence & 1: 40 & 1: 47 & 1: 56  & 1: 62 & 1: 64 & 1: 57 & 1: 47 & 1: 57 & 1: 64 \\
      & 0: 48 & 0: 46 & 0: 42  & 0: 45 & 0: 51 & 0: 58 & 0: 68 & 0: 56 & 0: 49\\
\hline
\end{tabular}
\caption{Results for observations. 1 / 0 denote the number of ARs in which the model detects / fails to detect the flow/magnetic field feature.}
\label{table_classified_observation}
\end{table}

\section{Absence of strong correlation between magnetic fields and flows}
\begin{figure}[!htb]
    \centering
    \includegraphics[scale=0.55]{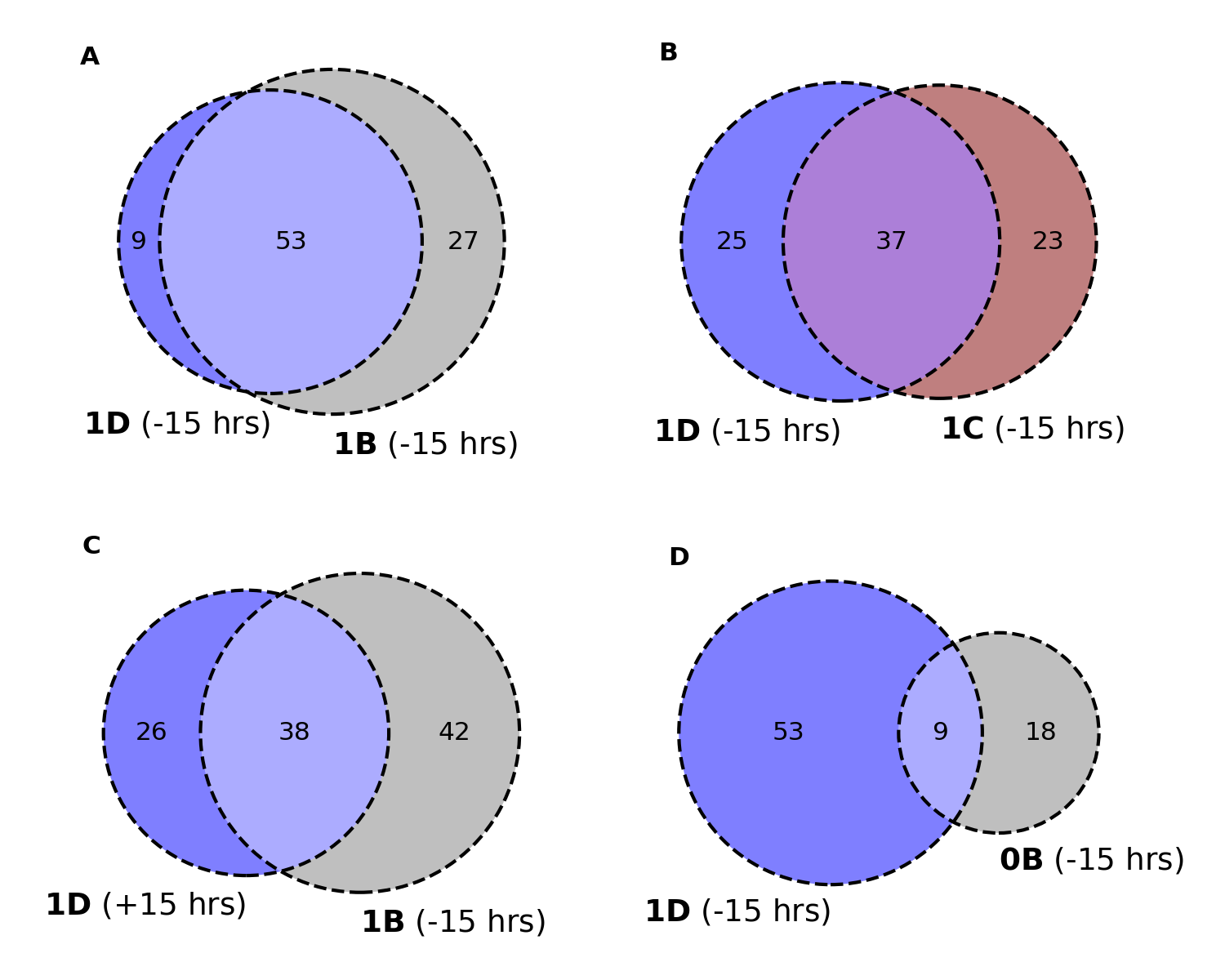}
    \caption{Venn diagrams depicting correlation between flow and magnetic field features in observations. Below each venn circle in a panel the feature and the machine prediction is indicated - for instance 1D (-15 hrs.) means ARs for which machine predicted as containing inflows in the interval -15 hrs.. The number of such ARs is given inside the venn circle.}
    \label{fig_venncombo}
\end{figure}

Prior studies have shown correlation between emerging active region flows and magnetic flux. As is shown in the Venn-diagram (see Figure~\ref{fig_venncombo}), we do not measure substantial correlations between flows and magnetic fields in the observations of active regions. Rather than imposing a flow-magnetic field correlation apriori, it is more objective to discover if any such connection exists through independent machines for these features. 

\bibliography{References}{}
\bibliographystyle{aasjournal}
\end{document}